\documentclass[12pt]{article}
\usepackage{graphicx}
\begin{document}
\title{Collapse Models and  Perceptual Processes}

\author{GianCarlo Ghirardi\footnote{e-mail:ghirardi@ictp.it}\\
{\small Emeritus, University of Trieste, and} \\ 
{\small  The Abdus Salam ICTP, Trieste, Italy}\\and\\
Raffaele Romano\footnote{rromano@iastate.edu}\\
{\small Department of Mathematics, Iowa State University, Ames, IA 50011 (USA)}}

\date{}
\maketitle

\begin{abstract}
Theories including a collapse mechanism have been presented various years ago. They are based on a modification of standard quantum mechanics in which nonlinear and stochastic terms are added to the evolution equation. Their principal merits derive from the fact that they are mathematically precise schemes  accounting,  on the basis of a unique universal dynamical principle, both for the quantum behavior of microscopic systems as well as for the reduction associated to measurement processes  and for the classical behavior of macroscopic objects. Since such theories qualify themselves not as new interpretations but as  modifications of the standard theory they can be, in principle, tested against quantum mechanics. Recently, various investigations identifying possible crucial test have been discussed. In spite of the extreme difficulty to perform such tests  it seems that recent technological developments allow at least to  put precise limits on the parameters characterizing  the modifications of the evolution equation. Here we will simply mention some of the recent investigations in this direction, while we will mainly concentrate our attention to the way in which collapse theories account for definite perceptual process. The  differences  between the case of reductions induced by perceptions and those related to measurement procedures by means of standard macroscopic devices will be discussed.  On this basis, we suggest a precise experimental test of collapse theories involving  conscious observers. We make plausible, by discussing in detail a toy model, that the modified dynamics can give rise to quite small but  systematic errors in the visual perceptual process. 
\end{abstract}

\section{A concise review of the measurement problem}
As is well known the paradigmatic case in which an embarrassing situation  emerges within a quantum scenario is the macro-objectification or measurement problem. It can be summarized by resorting to the idealized scheme proposed by von Neumann \cite{von}.

We consider a microscopic system $s_{micro}$, we assume that it has been prepared in the eigenstate $|\omega_{i}\rangle$ of an appropriate observable $\Omega$, and we suppose to be interested in determining  the value of this observable by a measurement process. We then consider a macroscopic apparatus $S_{macro}$ prepared in its ``ready" state $|S_{0}\rangle$ and we switch on a system-apparatus interaction inducing, in the time span in which the measurement takes place,  the process summarized by the second of the following equations:

\begin{eqnarray}
\Omega|\omega_{i}\rangle&=&\omega_{i}|\omega_{i}\rangle \\
|\omega_{i}\rangle\otimes|S_{0}\rangle&\rightarrow&|\omega_{i}\rangle\otimes|S_{i}\rangle,
\end{eqnarray}

\noindent where we have denoted as  $|S_{i}\rangle$ the orthogonal final states of the apparatus which are assumed to correspond to mutually exclusive perceptions of a conscious observer looking, e.g., to the ``pointer position" of the apparatus.

Then, if we trigger the apparatus with a state which is a superposition of different eigenvectors of $\Omega$, the linear nature of quantum mechanics implies the following evolution for the system-apparatus system:

\begin{equation}
[\sum_{i}c_{i}|\omega_{i}\rangle]\otimes|S_{0}\rangle\rightarrow \sum_{i}c_{i}|\omega_{i}\rangle\otimes|S_{i}\rangle.
\end{equation}

\noindent Equation (3) shows clearly that the system and the apparatus are now entangled, and, in particular, that the apparatus cannot be claimed to possess a precise macro property associated to our definite perceptions when  we look at it.\footnote{It has to be stressed that the unacceptable situation we have just mentioned does not derive from the fact that we have resorted to the (extremely) idealized description  of von Neumann of the measurement process. Actually, as it has been proved recently \cite{bassi}, the very requirement that one can ascertain a micro property with a reasonable degree of reliability together with the assumption of the universal linear character of the evolution, implies that superpositions of macroscopically and perceptively distinguishable states unavoidably occur. }

We consider it appropriate to close this section by the sharp statement by J. Bell:
\begin{quote}
{\it Nobody knows what quantum mechanics says exactly about any situation, for nobody knows where the boundary really is between wavy quantum systems and the world of particular events.}
\end{quote}

\section{Some proposed solutions which do not alter the quantum predictions}

We list and briefly comment here some of the most popular proposals to overcome the just mentioned problem, proposals which, however, fully agree with  quantum predictions for what concerns the outcomes of measurement processes. 

\begin{itemize}
\item  {\it Incompleteness.} The state is not everything. This way out of the macro-objectification problem considers the quantum specification of the state of a physical system,  given by the statevector, as not complete: to have an exhaustive description one has to add to the statevector further variables, whose knowledge would allow a more detailed specification of the objective situation of individual physical systems and would consent more accurate or even fully precise predictions of the outcomes of measurement processes. The paradigmatic example of this position is  the de Broglie-Bohm theory, which, in recent years, has seen an impressive set of mathematical investigations aimed to clarify the most delicate points of the theory\footnote{For an exhaustve and lucid discussn of the De Broglie Bohm theory we refer the reader to the book \cite{durr}}. The ensuing formal scheme is completely satisfactory, it represents a deterministic completion of the theory based on the addition to the state vector of the so called ``hidden variables", which are identified with the positions of all particles of the system, and  which are, in principle, unaccessible. The theory is logically consistent and, in our opinion, it represents an extremely interesting theoretical scheme which overcomes the  foundational problems of the theory.
\item{\it Limiting observability or invoking deoherence.} This perspective has many facets, as one can see by taking into account the positions of Jauch \cite{jauch}, Daneri {\it et al} \cite{daneri}, Joos and Zeh \cite{joos} and Zurek \cite{zurek}, among others. The philosophy which characterizes it is extremely simple: either in principle (Jauch, Daneri) or for practical reasons (Joos \&  Zeh, Zurek) one cannot find a way to put into evidence  that superpositions of macroscopically and perceptibly different situations actually occur. There are two fundamental objections which can be raised against this position: one derives from remarking that it never happened, in the history of science, that a theory leads to a nonsensical (in the sense of not matching our perceptions) situation - the superposition of perceptibly  different states - but, at the same time, that resorting to an approximation  one gets an acceptable picture of natural phenomena. Secondly, as remarked, among many others, by S. Adler \cite{adler}, {\it the decoherence approach has not solved the measurement problem}, this statement following from the well known fact that the correspondence {\it physical ensembles} $\rightarrow$ {\it statistical operators} is infinitely many to one in quantum mechanics, so that, as recognized even by the strongest supporters of the decoherence approach \cite{joos}: 
\begin{quote}
{\it no unitary treatment of the time dependence can explain while only one of these dynamically independent components is experienced.}
\end{quote}
According to these authors the emergence of our definite perceptions {\it might depend on our local way of perceiving}, a totally unacceptable position, in our opinion, due to the central role which it attributes to the conscious observer.
\item {\it Enriching Reality.} We have decided to use this term to identify both the Many Worlds \cite{everett, dewitt}  as well as the Many Minds \cite{albert1} interpretations. In the first case, it is assumed that when one reaches the macro level and  superpositions of macroscopically and perceptibly different states occur, the universe multifurcates, so that there are many universes in which all different macro situations become actual,  one of them for each different universe. The second attitude corresponds to assuming the same not for the actual occurrence of different  situations, but for the occurrence of all potentially possible
 perceptions: our brains are  built in such a way that there are different ``sheets" registering different perceptions, the different sheets of different observers having to be  synchronized among themselves in such a way that no intersubjective disagreement might occur.
\end{itemize}

At this point we consider it interesting to present the completely different position of the adherents to the collapse view of natural processes.

\section{Collapse Theories}

The central idea of this approach consists in contemplating that the linear and deterministic evolution of the standard theory has not a universal validity: the basic Scr\"{o}dinger's equation must be modified by the addition of nonlinear and stochastic terms, which  account for the definite  features of the reduction process.

As it is obvious, and as it has been stressed by many scientists,  macro-objects are characterized by the fact that they correspond to perceptually different locations of (some) of their macroscopic parts (typically their ``pointer"), so that it is quite natural to tackle ``the preferred basis problems" by assuming that the modified dynamics strives to make precise the positions of  physical objects. Given this, we can be fully precise about the features of collapse models. For simplicity we will make reference to the original proposal of ref.\cite{grw}.

\begin{itemize}
\item {\it States}. A Hilbert space $\cal{H}$ is associated to any physical system and the state of the system at time $t$ is represented by a normalized vector $|\psi_{t}\rangle$ of $\cal{H}$.
\item {\it Dynamics}. The evolution of the system is governed by Schr\"{o}dinger's equation. In addition, at random times, with a Poissonian distribution with mean frequency $\lambda$ (which is assumed to be proportional to the mass of the particle under consideration), each particle of any system is subjected to a spontaneous localization process of the form:
\begin{equation}
|\psi_{t}\rangle\rightarrow\frac{L_{n}(\bf x)|\psi_{t}\rangle}{||L_{n}({\bf x})|\psi_{t}\rangle||};\;\;\;L_{n}({\bf x})=(\frac{\alpha}{\pi})^{3/4} \exp[-\frac{\alpha}{2}({\bf\hat{x}}_{n}-{\bf x})^{2}].
\end{equation}
\noindent In this equation ${\bf\hat{x}}_{n}$ is the position operator of the $n-th$ particle of the system.
\item{\it Collapse probability}. One assumes that the probability density that the collapse for the $n-th$ particle occurs at the space point ${\bf x}$ is given by:
\begin{equation}
p_{n}({\bf x})=||L_{n}({\bf x})|\psi_{t}||^{2}.
\end{equation}
\item {\it Ontology}. Let $\psi_{t}({\bf x}_{1}, .....{\bf x}_{N})$ be the wave function in configuration space. Then:
\begin{equation}
m({\bf x},t)\equiv\sum_{n=1}^{N}m_{n}\int d^{3}{\bf x}_{1}...d^{3}{\bf x}_{1N} \delta^{(3)}({\bf x}_{n}-{\bf x})|\psi_{t}({\bf x}_{1}, .....{\bf x}_{N})|^{2},
\end{equation}
\noindent is assumed to describe the {\it density of mass} distribution of the system of the N particles under consideration in three-dimensional space as a function of time.
\item {\it The trigger mechanism}. The nicest feature of the model consists in the fact that the localizations increase with the number of particles. In particular, it can be rigorously proved that any localization of any particle implies a localization of the centre of mass of the whole system. Accordingly, in the case of an almost rigid body, its position suffers a localization with a frequency amplified, with respect to the one of the individual constituents, by the number of the constituents of the body.
\item{\it Choosing the parameters of the theory}. The original choice \cite{grw} of the values of the localization accuracy and of the mean frequency of the localization {\it for a nucleon} has been:
\begin{equation}
\lambda\simeq 10^{-16}sec^{-1},\;\;\; \alpha\simeq 10^{10} cm^{-2}.
\end{equation}
\noindent Note that with these choices a microscopic system suffers a localization about every $10^{7}$years, while a macroscopic system ($N\simeq 10^{23}$) one about every $10^{-7}$ sec. This is why the standard theory is left practically unchanged at the micro-level, while superpositions of macroscopically distinct states are suppressed in extremely short times.

\end{itemize}

Some important remarks:
\begin{itemize}
\item The physics depends essentially only on the product $\alpha\lambda$ with the only proviso that the localization accuracy must be much larger than the atomic dimensions in order that the modified dynamics leaves practically unaffected the internal motion.
\item Changing the above product of some orders of magnitude contradicts well established facts or it requires some important modification like the introduction of an appropriate cut-off.
\item The theory qualifies itself as a rival theory with respect to quantum mechanics and can be subjected to crucial tests with respect to this theory. Moreover, if one takes it seriously, one can get from it  indications concerning where to look for an hypothetical breaking of the superposition principle.
\item Recently a lot of attention has been paid to the possibility of devising crucial experimental tests of the theory against quantum mechanics. These tests cover a wide range of experimental situations. We will mention those making reference to the fact that the collapse processes tend to destroy quantum interference (the problem has been discussed in ref.\cite {arndt, hackermuller, gerlich}). Other interesting effects follow  from the fact that the collapse models imply spontaneous photon emission (see ref.\cite {fu, adler3, adler4}), as well as from the fact that energy is not conserved (see ref. \cite{grw, adler2}). For a general review of the problem of testing collapse models we refer the reader to \cite{adler5}.
\end{itemize}

\section{The problem of definite perceptions within collapse models}

In 1989, Albert and Vaidman \cite{albert2} have raised an important issue concerning the implications of collapse models with respect to the perceptual process itself. The challenge can be easily formulated: one has a neutral atom of spin $1/2$ in the spin state corresponding to $\sigma_{x}=+1$, which traverses a region in which an inhomogeneous magnetic field (along the $z$-direction, orthogonal to $x$ and to the propagation direction $y$) generated by a Stern-Gerlach device is present. As we all know, quantum mechanics predicts that the state of the atom, after the region of the apparatus has been traversed, is the equal weights superposition of its being {\it deflected upwards} and {\it deflected downwards}. On the  considered paths one puts a fluorescent screen with the following properties: when the screen is hit by the impinging atom, a relatively small number of its atoms (of the order of ten) are excited, the excited states having an extremely short life time, so, immediately after the hitting, we have a superposition of two states $|10\; photons\; from\; A\rangle$ and $|10\; photons\; from\; B\rangle$, a notation to denote that the photons originate, in the first case, from a point $A$, and in the second case, from a point $B$ of the screen. The experimental set up is depicted in Fig.[1]. 
\begin{figure}
\begin{center}
\includegraphics[scale=0.5]{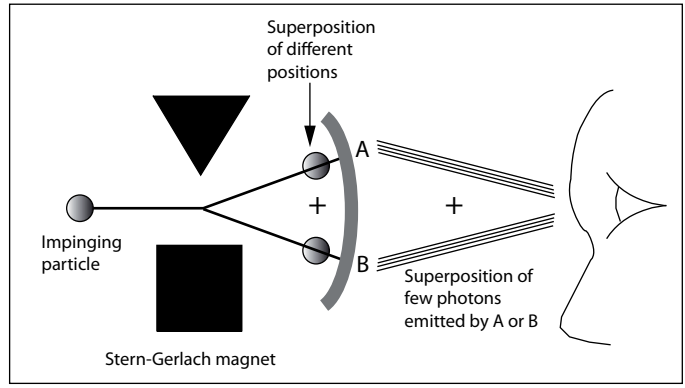}
\end{center}
\caption{The Albert and Vaidman proposal}
\end{figure}

The argument goes then as follows:
\begin{itemize}
\item In the time interval  before the photons reach the eye, the collapse dynamics, due to the small number of particles involved and due to the fact that collapses do not involve photons, is not able to induce a collapse, so that we actually have the linear superposition $ (1/\sqrt{2})[|10\; photons\; from\; A>+|10\; photons\; from\; B>].$ We assume that one can test, by an interference experiment, that the superposition is actually there.
\item Now we call into play a perceiving observer who looks at the screen to check whether he sees a spot from the point $A$ or from the point $B$. Since the threshold for visual perceptions is of the order of 5-7 photons, one is led to conclude that the observer will have a definite perception concerning the point in which there is a luminous spot. 
\item The conclusion follows: in the considered situation it is the conscious act of perception of the observer which determines the actual outcome, i.e. it induces the ``reduction" of the ``superposition" of two different and incompatible perceptions to only one of them. 
\end{itemize}

This point of view has been subsequently stressed by Albert himself \cite{albert3} in his interesting book: {\it Quantum Mechanics and Experience}:
\begin{quote}
{\it Suppose we want to stick with the GRW theory anyway. What would that entail? ... Well, we would have to deny that the measurement described above is over even once a recording exists. We would have to insist (and certainly this is an ineluctable fact, when you come right down to it) that no measurement is absolutely over, no measurement absolutely requires an outcome, until there is a ``sentient observer" who is actually aware of that outcome}.
\end{quote}

The inappropriateness of this statement should be clear to anybody. There is no doubt that, within collapse theories, any (non human)  detector aimed to check whether there is a spot at A or there a spot at B (e.g.,  the firing of one of two Geiger counters located along the  paths of the emitted photons, or even an apparatus such that its triggering by the superposition would put it  - in the quantum description - in the superposition of having its pointer pointing to ``From A" or ``From B"), would imply a reduction of the state. No conscious observer is necessary in collapse models. However, the argument of ref.\cite{albert2} undoubtedly represents an extremely serious and important challenge to collapse theories, a challenge  we accepted to face.

Shortly after the appearance of Ref.\cite{albert2}, Aicardi, Borsellino, Ghirardi and Grassi  analyzed \cite{aicardi} the perceptual process adopting precisely  the point of view of collapse theories \footnote{Borsellino has been the first and most prominent representative of neurophysicists in Italy.} . 

The perceptual process involves the following steps:

\begin{itemize}
\item Transmission of the stimulus from the rods of the retina to the lateral geniculate body and from this to the higher visual cortex, the transmission of the nervous signals taking place along the axons connecting the various regions.
\item The transmission mechanism implies the passage of Na and K ions from the internal to the external region of the axon through the Ranvier nodes. The axon is coated by the myelin sheet, which is precisely $10^{-5} cm$ thick (the characteristic reduction distance of the collapse models).
\item If one takes seriously this line and assumes a very prudent attitude about the number of particles involved, it turns out that the process involves, {\it just within perceptual times}, the displacement of a sufficient number of particles in order that the collapse mechanism become effective: one of the two nervous signals is suppressed and  a definite perception emerges.
\end{itemize}

\section{Have  the parameters of collapse theories been appropriately chosen?}

As we have discussed in the previous section, if one makes the choices (7) for the parameters of the model under considerations, one can show that macroscopic measurement processes have outcomes and conscious observers have definite perceptions, even when the perceptual process is triggered by a genuine quantum superposition and no apparatus enters into play. However, as we have just mentioned, in  ref.\cite{aicardi}, in order to account for conscious perceptions, the whole process of nervous transmission in the brain has  been taken into account. This fact does not seem fully satisfactory to some authors, and has been challenged, e.g., by S. Adler \cite{adler2}. He perfectly agrees on the fact that collapse theories, when few particles are involved, turn out to make predictions in full agreement with those of standard quantum mechanics, as well as on the fact that one must require that when one reaches the macroscopic  level, linear superpositions must be suppressed. However, there is a mesoscopic region (from the point of view of the number of involved particles) about which there might be some disagreement. Typically, as already indicated, S. Adler is not fully satisfied with the fact that the whole process of nervous transmission is necessary in order  that the collapse takes place, and suggests that reduction must already take place when the rods of the eye register the signal.

Another fact should be mentioned. In the formulation we have adopted, collapse theories are not able to consistently deal with systems containing identical constituents, because collapses do not respect the (anti)symmetry requirements for such constituents. This requires a modification of the formalism and a quite nice reformulation of the original theory which solves this problem has been worked out by Pearle \cite{pearle} and by Pearle {\it et al.} \cite{GPR}. However, taking into account the identity of the elementary constituents, the efficiency of the trigger mechanism is increased\footnote{We refer the reader to Ref.\cite {bassi} for a detailed discussion of this problem.}. Without entering into the details of the mathematical analysis we summarize the general situation: the localization accuracy $1/\sqrt{\alpha}$ cannot be decreased without inducing unacceptable effects, so, the game we can play is entirely related to possible changes of the mean frequency $\lambda$ of the localizations. The ensuing situation is the following:

\begin{itemize}
\item The choice for $\lambda$ made in the original collapse model \cite{grw} leads to the classical behavior of system with more than $10^{13}$ particles, while there is a continuous transition  to a complete quantum behaviour for numbers appreciably smaller than this. In particular when only $10^{4}-10^{5}$ particles are involved, the behaviour is certainly still quantum.
\item On the other hand, if one pretends, with Adler, the collapse to occur at the level of rods, in which case the displacement of a number of particles of the just indicated order is involved, one must appreciably increase the value of $\lambda$. As discussed by Adler a choice $10^{8}$ times greater than the original one is perfectly able to induce the collapse in the time in which the rearrangement of the constituents in the rod takes place. Accordingly, Adler suggest to choose for $\lambda$ the value $\lambda\simeq 10^{-8\pm 2}sec^{-1}$. This request, however, implies a too fast increase of the energy of the universe due to collapses and, as a consequence, it requires some modifications  which alter the original simplicity of the theory. 
\end{itemize}

At any rate, the present analysis is useful to point out an essential fact which distinguishes the case of collapses due, typically, to superpositions of different positions of a macroscopic body from those taking place when our apparatus is triggered by a micro superposition whose states induce directly different perceptions. We discuss this point in the next section. 

\section{A comparison of the characteristic times of the hamiltonian evolution and of the reduction processes}

Just to give an idea of the situation, let us take first of all into account a free macroscopic system, like a pointer of $1 gr$, initially in a state with a spread of the c.o.m.  position of the order of $10^{-5}$ cm. If we want to get an idea of the changes induced in the statevector by the hamiltonian evolution, a good estimate is given by the time it takes to double its spread, or (considering a physically more meaningful spread) to reach a spread of $10^{-1}$ cm, which we  assume to be  smaller than the separation of the pointer positions related to the different outcomes in the measurement. This computation leads to the conclusion that, during the time in which the hamiltonian evolution changes appreciably the state of the object, it suffers, according to the original choice of the parameters, from $10^{25}$ to $10^{29}$ localizations. In the case considered by Adler, these numbers have to be amplified of a factor $10^{8}$.

On the contrary, for what concerns the perceptual process, when the perceptual apparatus is triggered by a superposition of microscopically different states (photons from A or from B), both in the original as well as in Adler's case, the reduction takes  place just in the relevant times, i.e. the perceptual ones, during which the statevector changes from the superposition to one of its terms. And here a new feature emerges. Important differences in the dynamics occur when the two processes, the hamiltonian evolution and the changes induced by the localizations, become competitive, with respect to the case in which the localizations are enormously more frequent that the time required by hamiltonian changes. We will make clear this point by discussing, in the next section, a specific toy model (which has nothing to do with perceptions) in which we have a competition of the hamiltonian evolution and the reduction process. The study of this case is interesting because it allows  to make plausible that collapse models might have consequences on the unfolding of the perceptual process.

Actually, the arguments we will present have already been outlined in a paper \cite{gperc} by one of us. However, the treatment of ref.\cite{gperc} was rather qualitative. The real novelty consists in the fact that here we will present rigorous results supporting the proposal.

\section{A toy model making plausible that collapse models can be tested in perceptual experiments}

In this section, motivated by the considerations of the previous one, we will discuss in detail a toy model, characterized by an hamiltonian evolution and by a collapse process which can be considered as competing. The aim is to check whether under such conditions some interesting effects might arise. The toy model we will consider has nothing to do with actual processes triggering different perceptions, and is a trivial two dimensional example, the two possible states in which the collapse strives to drive the statevector mimicking, in a sense, the two possible perceptions in an experiment of the kind presented in ref.\cite{albert2}. 

Here is the model. We have a spin $1/2$ particle evolving according to  Schr\"{o}dinger's equation with the hamiltonian:
\begin{equation}
H=\hbar\omega \left( \begin{array} {cc}
0 & 1 \\ 1 & 0 \end{array} \right),
\end{equation}
and subjected, with mean frequency $\lambda$, to reduction processes on the eigenstates of $\sigma_{z}$. We assume that the initial state of the system is:

\begin{equation}
\psi(0)= \left( \begin{array} {c} a\\ ib \end{array} \right),
\end{equation}
where $a$ and $b$ are real numbers that we choose to take the values\footnote{We have made $a\neq b$ because the  effects we are going to analyze depend on the initial state and the choice we have made turns out to be particularly appropriate for the emergence of the discrepancies we are interested in.}
\begin{equation}
a=\sqrt{0.48},\;\; b=\sqrt{0.52}.
\end{equation}

Let us write the statistical operator at time $t$ in the following form:

\begin{equation}
\rho(t)= \left( \begin{array}{cc} 
\rho_{1}(t) & \rho_{3}(t) \\ \rho^{*}_{3}(t) & (1-\rho_{1}(t)) \end{array}  \right),
\end{equation}
and let us call:

\begin{equation}
P_{+}= \left( \begin{array}{cc} 1 & 0 \\ 0 & 0 \end{array}\right),\;P_{-}= \left( \begin{array}{cc} 0 & 1 \\ 0 & 0 \end{array}\right),
\end{equation}
the projection operators on the eigenstates of $\sigma_{z}$ belonging to the eigenvalues $+1$ and $-1$, respectively.

The quantum dynamical semigroup equation for the statistical operator is then:
\begin{equation}
\frac{d\rho(t)}{dt}= -\frac{i}{\hbar}[H,\rho(t)]+\lambda P_{+}\rho(t) P_{+}+ \lambda P_{-}\rho(t) P_{-}-\lambda\rho(t).
\end{equation}

Note that the  time which is necessary in order that the quantum evolution induces appreciable changes of the statevector (e.g. the time in which the spin makes a complete rotation around the $z$-axis) is $\tau=\pi/2\omega$, while the characteristic time between two reduction processes is $1/\lambda$. Since we want  the reduction to occur within times of the order of the perceptual times, we will choose $\lambda=10^{2} sec^{-1}$. The parameter determining the ratio of the inverses of the times of the unitary evolution and of the reductions will be denoted as $\epsilon=\omega/\lambda$. 

With reference to the expression (11) of the statistical operator we call the attention of the reader on the fact that $\rho_{1}(t)$ gives the probability that, at time $t$, the spin is in the state corresponding to $\sigma_{z}=+1$, while the vanishing of $\rho_{3}(t)$ tells us that reduction has actually taken place. In fact, in the present model it is obvious that reductions take place at the individual level, so that, here, it is legitimate to study the collapse process looking only at the statistical operator.
 
We  analyze now in detail the evolution of the statistical operator for 2 quite different values of $\epsilon$. First of all, to mimik the typical situation of realistic collapse models,  we  make the reductions much more frequent than the time in which the hamiltonian changes appreciably the statevector by choosing $\epsilon=10^{-6}$, so that one million reductions occur during the time in which a rotation of $2\pi$ of the spin  is caused by the hamiltonian. Alternatively we will consider the case in which $\epsilon=10^{-2}$, corresponding to a much smaller number of reductions $(\simeq 100)$ during the same time.

\subsection{A remark concerning the asymptotic behavior}

In realistic collapse theories and in the case of a macroscopic body in a superposition of macroscopically different states the time evolution goes as follows:
\begin{itemize}
\item In times of the order of $\lambda_{macro}^{-1}\simeq 10^{-7} sec$ the reduction process takes actually place leading to one of the two superposed states with the correct quantum probabilities.
\item Subsequently, the statevector remains, for an incredibly long time (many times the age of the universe), the one to which the reduction process has led.
\item However, asymptotically, i.e. for $t\rightarrow +\infty$, the statevector changes and it does not tend to a precise limit. This is a mathematical fact which has no physical relevance.
\end{itemize}

The situation is rather different in the case we are discussing, i.e. when one has an equation of the quantum dynamical semigroup type  which admits a steady solution. In such a case a completely general theorem asserts that any solution must tend to it when $t\rightarrow\infty$. Our model manifestly exhibits the steady solution  $\rho(t)=(1/2)I$. This implies that the unfolding of the  evolution process takes place in this way:

\begin{itemize}
\item The reduction time is characterized essentially, as before, by $\lambda^{-1}$.
\item Subsequently, reduction takes place to one of the superposed states with a probability depending on the initial state and it lasts for a certain period.
\item However, at a certain time a new regime occurs, since, as we have stressed, the statistical operator must take the form $\rho(t)=(1/2)I$. It is important to have clear that what really matters is the second step, i.e. the detailed features of the reduction process and the time for which it lasts. In fact,  here we are dealing with a toy example, mimicking only a very small part of the real process. Actually, if reduction takes place and lasts even for a short (but long with respect to the perceptual times) period, the very reduction process, i.e. the actual emergence of a definite perception, triggers  other changes involving an extremely large number of particles. For instance other parts of  the  brain are surely involved besides the axons which have transmitted the signal, or, just to make a trivial example, we might ask the person whose eyes have been stimulated by the superposition to write on a piece of paper either $A$ or $B$ according to region from which he or she has seen the luminous spot. This fact has to be taken present because it has an important consequence: after the perception emerges other macroscopic changes are implied by its specific character (having perceived a spot from $A$ or from $B$) and such changes are then frozen by the collapse mechanism which is active for all particles of the universe. So, also in the present case, after the collapse corresponding to the emergence of a definite perception, the reduction lasts practically forever. 
\end{itemize}

\begin{figure}
\begin{center}
\includegraphics[scale=1.2]{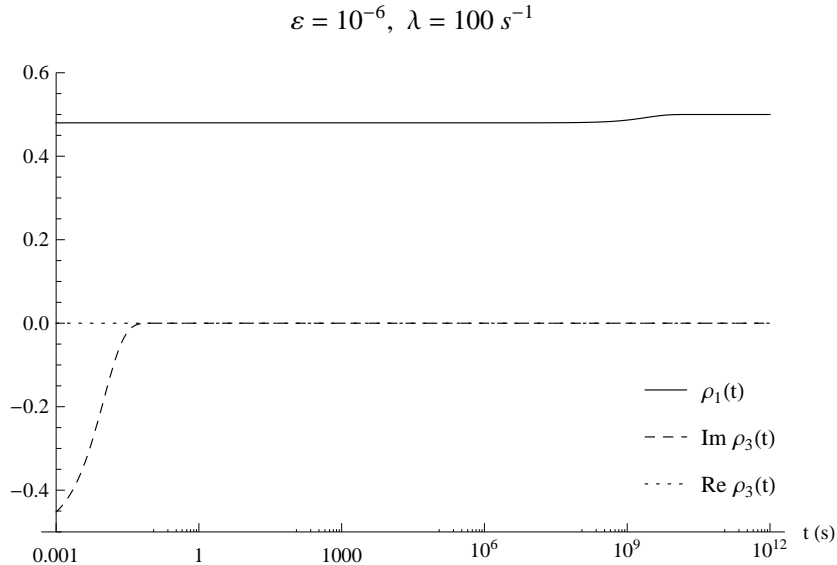}
\end{center}
\caption{The behavior of the statistical operator in the case $\epsilon=10^{-6}.$}
\end{figure}

\section{The actual unfolding of the process of Section 7}

We now present the results of the detailed analysis we have performed of the time behavior of the statistical operator for the two considered choices of $\epsilon$. We summarize our results in a series of figures in which we have plotted $\rho_{1}(t)$ and $\rho_{3}(t)$ of Eq.(11) under the dynamics (13).
We start, in Fig.2, with the case in which  $\epsilon$ is very small ($10^{-6}$), like in actual collapse processes involving macroscopic systems.
The figure shows clearly (the dashed and the dotted lines) that in a time of few hundredth of a second reduction takes place and leads to the eigenstate $\sigma_{z}=+1$ with the correct quantum probability $a^{2}=0.48$ implied by the form (9) of the initial state. The reduced state remains then unaltered for a quite long time ($\simeq 10^{9} sec$). Subsequently the asymptotic behaviour,  i.e. the one corresponding to a probability of 50\% for the reduction to one  of the eigenstates $\sigma_{z}=\pm1$, emerges.

\begin{figure}
\begin{center}
\includegraphics[scale=1.2]{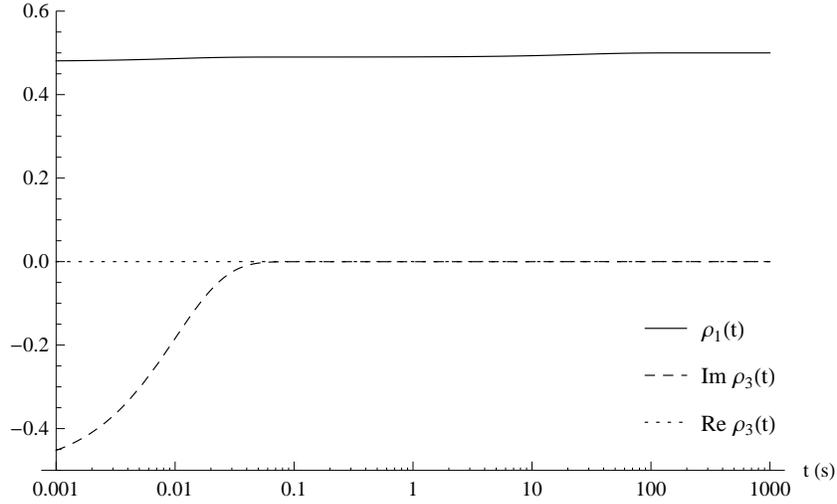}
\end{center}
\caption{The behavior of the statistical operator in the case $\epsilon=10^{-2}.$}
\end{figure}

We pass now to discuss the most interesting case, i.e. the one in which $\epsilon$ is appreciably increased ($\epsilon=10^{-2}$). Figure 3 (pay attention to the change of scale of the abscissa) looks similar to the previous one but there are some  differences. They are not evident from Fig. 3 since use has been made of an inappropriate vertical scale. To put into evidence these features, their relevance and to discuss their physical implications one has to amplify the vertical scale. This has been done in Fig.4 which shows various relevant facts. First of all, after the reduction has taken place (in a time of the order of $10^{-2}$ sec, as shown in Fig.3), the function $\rho_{1}(t)$ exhibits a plateaux, lasting for about 6-7 seconds, before the asymptotic regime emerges. The noticeable fact is that in this time interval (which, as discussed before is sufficient for inducing other macroscopic changes which are then frozen by the theory)  the probability of the outcome $\sigma_{z}=+1$ amounts to more than 49\%, i.e. it differs from the one of 48\% required by quantum mechanics. It seems that the fact that the dynamical set up of the perceptions and the reductions due to the collapses become competitive can lead to systematic perceptual discrepancies from the quantum predictions.

\begin{figure}
\begin{center}
\includegraphics[scale=1.2]{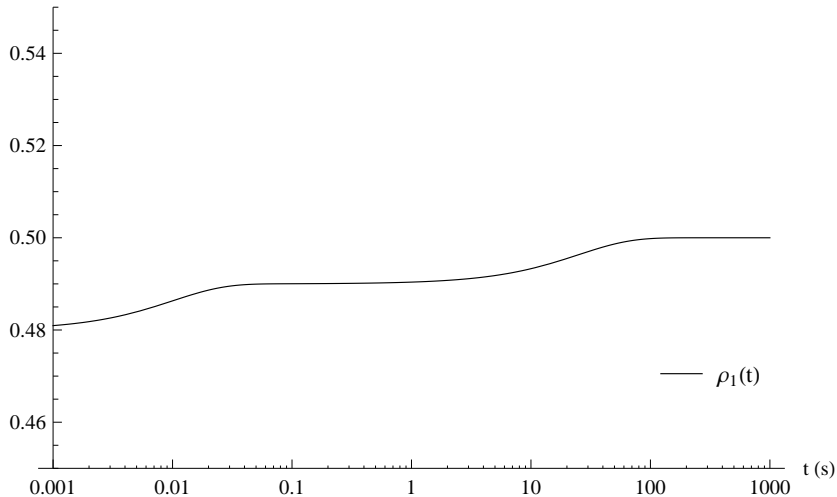}
\end{center}
\caption{The vertically amplified behavior of the function $\rho_{1}(t)$  in the case $\epsilon=10^{-2}.$}
\end{figure}

\section{A summary of the situation}
Our analysis gives some clear indications about the perceptual process. If one assumes that  in the case in which one triggers the visual apparatus with states corresponding to definite and different perceptions, randomly distributed with probabilities 48\% and 52\% between ``a spot at A" and ``a spot at B" , no perceptual errors occur,  the person subjected to the test would give  answers reproducing the considered distributions (actually one could also check that in each individual process the perception matches the stimulus). On the other hand, in the case in which one triggers the perceptual apparatus with a state which is a superposition of the just mentioned states which should yield, according to standard quantum theory, once more to  a 48\%-52\% distribution, some systematic errors seems to emerge,  implying a violation of the quantum predictions. In a sense, one might pictorially describe the situation by claiming that in the second case, the brain (or the rods) has (have) to make the additional work of ``reducing the state" and this might affect systematically the occurrence of the outcomes. Our analysis gives some (extremely) vague support to this idea and suggests that among the devised experiments to test collapse models versus the standard theory, also experiments involving perceptual process might and should be performed.

\section{Rejecting a possible criticism}

There is an  aspect of the dynamical reduction mechanism which deserves further investigations. We have been led to the idea that when directly stimulated by a superposition, the perceptual apparatus might yield distribution of outcomes (very slightly) disagreeing with those implied by quantum mechanics. However, if the discrepancies can be actually put into evidence,  we must check that the newly suggested peculiarity of the collapse models does not lead to consequences allowing faster than light effects. In fact, suppose the same effect, i.e. the enhancement of perceptions favoring ``a spot from A",  of triggering the perceptual apparatus with the state we have considered, ($a|10\; photons\; from\; A\rangle+b|10\; photons\; from\; B\rangle$) occurs also in the case when one triggers the apparatus with the orthogonal state $a|10\; photons\; from\; A\rangle-b|10\; photons\; from\; B\rangle$, while no perceptual mistake occur when triggering the apparatus with a state corresponding to a definite perception (i.e. one of the two of the superposition). In such a case one might entangle the spin states of our particle with states of a further physical system  far apart from the region in which the photons beams are, in the following way:
\begin{equation}
|\Psi>=\frac{1}{\sqrt{2}}\{|\chi_{1}\rangle\otimes[a|A>+b|B>]+|\chi_{2}\rangle\otimes[a|A>-b|B>],
\end{equation}

\begin{figure}
\begin{center}
\includegraphics[scale=1.2]{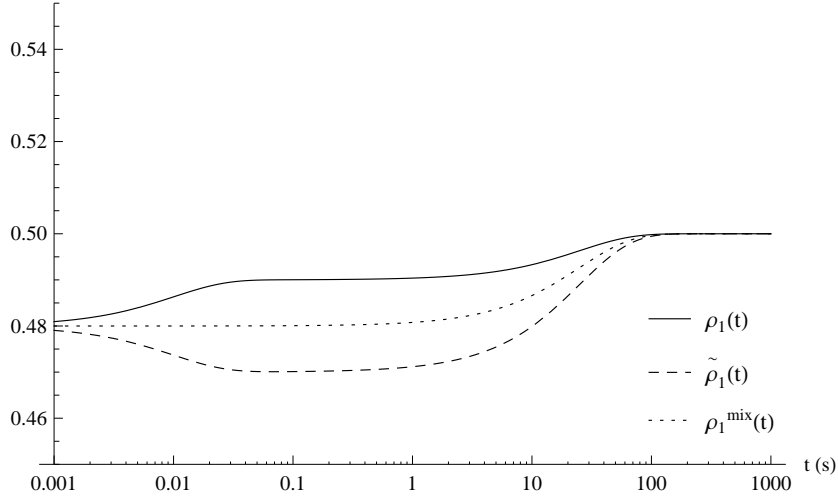}
\end{center}
\caption{The proof that no faster than light effect can occur}
\end{figure}

\noindent where we have used the notation  $|A>$  for the state $|10\; photons\; from\; A\rangle$ and $|B>$  for the state $|10\; photons\; from\; B\rangle$, respectively. In this case, the guy, let us call him Bob, who has access to the system whose states are $|\chi_{1}>$ and  $|\chi_{2}>$ can choose whether to measure an observable which has such states as eigenstates, leading then to reduction to one of the states $[a|A>\pm|B>]$. Alternatively, Bob might choose to measure  an observable having as eigenstates $[|\chi_{1}>\pm|\chi_{2}>]$ inducing then a reduction to either the state $|A>$ or to the state $|B>$. Then, the far away observer (Alice) would have his/her perceptual apparatus triggered either by the states corresponding to definite perceptions or by their superposition. However, if in the second case (reduction   to any one of the states $[a|A>\pm|B>]$) the perceptual process of Alice would be affected in the same way,  Alice would perceive more outcomes corresponding to $|A>$ with respect to those implied by quantum mechanics and she might become aware (in the long run) of the kind of measurements that Bob has decided to perform. Obviously, to avoid faster than light effects, we must guarantee  that this cannot happen. The request implies that the perceptual errors associated to triggering Alice visual apparatus with the state in which the sign of $b$ is changed should compensate those for which no change has been performed. We know that things must go in this way because there is a general proof \cite{bell} that  collapse theories do not imply superluminal effects. We have considered it useful to check this fact also with reference to our toy model. In Fig.5 we have plotted the first term of the statistical operator  for three different initial conditions:

\begin{enumerate}
\item $\rho_{1}(t)$: the initial state is $[\sqrt{0.48}|10\; photons\; from\; A\rangle+i \sqrt{0.52}|10\; photons\; from\; B\rangle]$,
\item $\tilde{ \rho}_{1}(t)$: the initial state is $[\sqrt{0.48}|10\; photons\; from\; A\rangle- i\sqrt{0.52}|10\; photons\; from\; B\rangle],$
\item $\rho_{1}^{mix}(t)$: the initial state is the statistical mixture, with weights 0.48 and 0.52 of the states corresponding to definite perceptions associated to $|10\; photons\; from\; A\rangle$ and $|10\; photons\; from\; B\rangle.$
\end{enumerate}
 As it is evident, the function for the third case turns out to be precisely the average of the first two cases. No superluminal signal can occur.

\section{Conclusions}
We have made plausible that, besides the various possible tests of collapse models versus quantum mechanics, also resorting to specific processes involving the direct triggering of the perceptual apparatus by a quantum superposition of microscopic states might help in discriminating between these two descriptions of natural processes.

\section*{references}

\smallskip

\end{document}